\documentclass[10pt,preprint,showpacs,twocolums]{revtex4-1}
\usepackage{amsfonts}
\usepackage{amsmath,amssymb}
\DeclareMathOperator{\sech}{sech}
\usepackage{graphicx}
\usepackage{subfigure}
\usepackage{bm}
\usepackage{epstopdf}
\begin{document}
\title{Superfluidity of a dipolar Fermi gas in 2D optical lattices bilayer}

\author{A. Camacho-Guardian and R. Paredes\footnote{Corresponding author: rosario@fisica.unam.mx}}                   

\affiliation{ 
 Instituto de F\'{\i}sica, Universidad
Nacional Aut\'onoma de M\'exico, Apartado Postal 20-364, M\'exico D.
F. 01000, Mexico. }

\pacs{67.85.-d, 03.75.Ss, 03.75.Gg}

\date{\today}
\begin{abstract}
We propose a model for addressing the superfluidity of two different Fermi species confined in a bilayer geometry of square optical lattices. The fermions are assumed to be molecules with interlayer $s-$wave interactions, whose dipole moments are oriented perpendicularly to the layers. Using functional integral techniques we investigate the BCS-like state induced in the bilayer at finite temperatures. In particular, we determine the critical temperature as a function of the coupling strength between molecules in different layers and of the interlayer spacing. By means of Ginzburg-Landau theory we calculate the superfluid density. We also study the dimerized BEC phase through the Berezinskii-Kosterlitz-Thouless transition, where the effective mass leads to identify the crossover from BCS to BEC regimes. The possibility of tuning the effective mass as a direct consequence of the lattice confinement, allows us to suggest a range of values of the interlayer spacing, which would enable observing this superfluidity within current experimental conditions.
\end{abstract}
\date{\today}
\maketitle

{\it Introduction.-}
\label{intro}
Although the underlying mechanism for the occurrence of high $Tc$ superconductivity has not been fully elucidated, it has been established that it must include two essential features, long range interactions among its constituents and transport in parallel layers \cite{Leggett,Orenstein}. At present, experiments carried out with ultracold molecules offer the possibility of emulating such conditions as a result of their precise control achieved in the laboratory \cite{Baranov,Carr,Lahaye, Krems,Trefzger,Mingwu}. Indeed, successful realizations of ultracold dipolar gases of molecules, with magnetic or electric nature, either with permanent or induced dipole moment have already been performed. Besides this interest, it has been proposed that efficient quantum information processing \cite{Demille} can be achieved with systems with long range interactions and movement confined in a 2D geometry. Similarly, the predicted Wigner crystalline phase \cite{Buchler} also shares those distinctive attributes.

Here, we address on the study of superfluidity in dipolar Fermi molecules confined in 2D optical lattices. For this purpose we consider the system described below and work within the mean-field perturbative analysis using the integral functional formalism to determine the critical transition temperature. The superfluid transition is analyzed in the Ginzburg-Landau (GL) scheme where the superfluid density is calculated. We found that assisted by the 2D lattice confinement the attractive interactions between molecules in different layers induces interlayer paring \cite{Pikovski, Potter, Zinner} of both types, BCS and bound molecular BEC states.  
   
\begin{figure}[htbp]
\begin{center}
\includegraphics[width=3.5in]{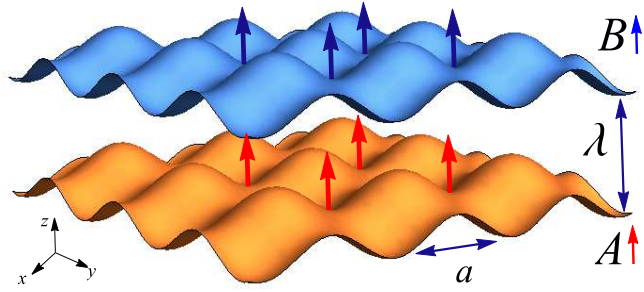} 
\end{center}
\caption{(Color online) Schematic representation of the dipolar Fermi gas. Polar molecules in the up (down) layer can be mapped into the specie labeled with $\uparrow$ ($\downarrow$) when the gas is described in a 2D layer. }
\label{Fig1}
\end{figure}

{\it Model.-} Our system consists of a gas of dipolar Fermi molecules placed in two parallel optical lattices in 2D. The configuration of the optical lattices in both layers has the same structure, a square lattice with constant lattice of size $a$. See Fig. \ref{Fig1}. In the presence of an electric field perpendicular to the layers, the dipoles are aligned along the same direction and, consequently, their interactions within the same layer are suppressed by intermolecular repulsion \cite{Baranov}. In contrast, dipoles in different layers attract each other at short range, while repelling each other at large distances. Thus, the interaction among dipoles has a specific form resulting from such a configuration. As schematically shown in Fig. \ref{Fig1}, such an array can be mapped into a system of fermions in two different hyperfine spin states, with interactions within an effective 2D environment. 
Fermions moving in up and down layers can be associated to two different hyperfine spin states labeled with $\uparrow$ and $\downarrow$ respectively. The interaction potential between fermions in different layers (or different hyperfine spin states) has the form \cite{Pikovski}

\begin{equation}
 V_{dip}(\vec{r})=d^2\frac{r^2-2\lambda^2}{(r^2+\lambda^2)^{5/2}},
\label{V_int}
\end{equation}
being $r$ the in-plane separation between fermions, $d$ the dipole moment and $\lambda$ the interlayer spacing. The Hamiltonian that describes such a system is 

\begin{equation}
 \hat{H}=\sum_{\alpha=A,B}\int d^2r\, \hat{\Psi}^{\dagger}_{\alpha}(\vec{r})H_{0}(\vec{r}) \hat{\Psi}_{\alpha}(\vec{r})+\frac{1}{2}\int \int d^2r\ d^2r'  \hat{\Psi}^{\dagger}_{A}(\vec{r}) \hat{\Psi}^{\dagger}_{B}({\vec{r}}^{\>\prime})V_{dip}({\vec r},{\vec r}^{\>\prime}) \hat{\Psi}_{B}(\vec{r}^{\>\prime}) \hat{\Psi}_{A}(\vec{r}),
\end{equation}
where $H_0(\vec{r})$ is the ideal term that includes the kinetic energy and the external potential created by the 2D optical lattice $V_{latt}(\vec{r})= V_0\left(\sin^2(x\pi/a)+\sin^2(y\pi/a)\right)$. The field operators $ \hat{\Psi}_{A}({\vec r})= \sum_{\vec k} \phi_{\vec k} ({\vec r})  \hat{a}_{\vec{k} }$ and $ \hat{\Psi}_{B}(\vec r^{\>\prime})=\sum_{\vec{k}} \phi_{\vec{k}} (\vec{r}^{\>\prime})  \hat{b}_{\vec{k} }$ satisfy the usual commutation relations for fermions. The in-plane energy dispersion of the ideal Fermi gas within the tight binding approximation is $\epsilon_{\vec{k}}=-2t(\cos{k_x a}+\cos{k_y a})$ being $t$ the hopping strength among nearest neighbors. It is worth mentioning that in real experiments, besides the optical lattice confinement, the atoms typically move under the influence of a harmonic confinement. However, since the frequency $\omega$ of this potential (due to the magnetic trapping) is so small, the curvature imposed on the  optical lattice can be neglected; that is, $\hbar \omega \ll t$. Therefore, the Hamiltonian of the two component Fermi gas in the momentum representation adopts the form

\begin{equation}
 \hat{H}=\sum_{\vec{k}}(\epsilon_{\vec{k}} - \mu) ( \hat{a}^{\dagger}_{\vec{k}} \hat{a}_{\vec{k}}+\hat{b}^{\dagger}_{\vec{k} }  \hat{b}_{\vec{k} })+\frac{1}{2\Omega} \sum_{\vec{k}, \vec{k'} ,{\vec q}} V_{dip}(\vec{k}-\vec{k}') \hat{a}^{\dagger}_{\vec{k}} \hat{b}^{\dagger}_{\vec q-\vec{k}} \hat{b}_{\vec{q}-\vec{k'}} \hat{a}_{\vec{k'}},
\label{Hk}
\end{equation}
where $V_{dip}(\vec{k}-\vec{k}')$ is the Fourier transform of $V_{dip}({\vec r}- {\vec r}^{\>\prime})$. In the last equation we have  introduced the chemical potential to account for the conservation of the total molecule number, and have already considered the lattice geometry. We note that $\Omega= N_x \times N_y$, being $N_x$ and $N_y$ the number of sites along $x$ and $y$ directions respectively.

To study the superfluidity in the model, we use functional integral techniques. The action for the ideal Hamiltonian term is,
\begin{equation}
S_0=\sum_{k}\phi^\dagger(k)\cdot\left[ i\hbar\omega_n \mathbb{I}- (\epsilon_{\vec{k}}-\mu)\sigma_z\right]\cdot\phi(k),
\end{equation}
where we use the abbreviations, $k=(\vec{k},\omega_n)$, $\phi=(\phi_{A}(k),\phi_B^*(k))^T$ with $\phi_{\alpha}$ Grassmann fermionic numbers, $\mathbb{I}$ is the identity in 2D, and the fermionic Matsubara frequencies are $\hbar\omega_n= (2n+1) \pi/ \beta$, with $\beta=1/k_B T$. $\sigma_\alpha$ ($\alpha=x,y,z$) are the Pauli matrices. To write the action associated to the interaction energy term, we introduce the auxiliary Hubbard-Stratonovich transformation in terms of the bosonic field $\Delta$

\begin{equation}
\langle\Delta (\vec{k})\rangle = -\sum_{\vec{k}'} V_{dip}(\vec{k}-\vec{k}')\langle\phi^\dagger_{A,-\vec{k}'} \phi^\dagger_{B,\vec{k}'}\rangle,
\end{equation}
that is, the action of such field creates a pair of particles with the same momentum ${\vec k}$ in different layers $A$ and $B$. In terms of the fields $\phi$ and $\Delta$ the action is 

\begin{equation}
S[\phi,\Delta]=\sum_{k,q}\Phi^\dagger(k,q)\cdot\left(i\hbar\omega_n \mathbb{I}-(\epsilon_{\vec{k}}-\mu)\sigma_z+\Delta(k,q)\sigma_x\right)\cdot\Phi(k,q)- \frac{1}{\Omega} \sum_{k,k',q} \Delta(q,k)V^{-1}_{dip}(\vec{k}-\vec{k}')\Delta(q,k'),
\end{equation}
being $\Phi(k,q)=(\phi_{A,k},\phi_{B,q-k})^T$.
Since this action has a quadratic form in the fields $\Phi$ one ends with the following expression for the effective action
\begin{equation}
S_{eff}[\Delta]= \sum_{{\vec k}, {q}}  \text{Tr} \ln \left( \mathbb{G}_{\Delta}^{-1} \right)- \frac{1}{\Omega} \sum_{k,k',q} \Delta(q,k)V^{-1}_{dip}(\vec{k}-\vec{k}')\Delta(q,k') ,
\label{Seff}
\end{equation}
with $\mathbb{G}_{\Delta}^{-1}=\left(i\hbar\omega_n \mathbb{I}-(\epsilon_{\vec{k}}-\mu)\sigma_z+\Delta(k,q)\sigma_x\right)$.
From this equation and within first order perturbation theory, the equation for the critical temperature, that signals the transition to the superfluid state, is 
\begin{equation}
\Delta_0({\vec k})=\sum_{ {\vec k'}} V_{dip}({\vec k}-{\vec k'}) \frac{ \Delta_0 ({\vec k'}) \tanh \left( \beta \xi_{\vec{k'}}/2 \right)} {2\xi_{\vec{k'}}},
\label{gap}
\end{equation}
where $\xi_{\vec{k}}=\sqrt{(\epsilon_{\vec k}-\mu)^2 + \Delta_0(\vec{k})^2}$. In this work we have restricted to the case $q=0$ and to an $s$-like gap. That is, assuming the momentum of the center of mass of the two-body collision term being zero, as in the original BCS scheme. 
 
{\it Critical Temperature.-} The physics of the system is completely determined by the dipolar interaction strength $a_d=m_{eff}d^2/\hbar^2$ and the dimensionless parameters $\Lambda= \lambda/a$ and $\chi=a_d/\lambda$, with $m_{eff}= \hbar^2/2ta^2$ as the effective mass. In principle, within our scheme, interlayer BCS pairs are formed for arbitrary values of these parameters \cite{Landau}. On the other side, consideration of the two-molecule collisions is in order, since, as in the homogeneous case \cite{Pikovski}, the two-molecule interlayer potential $V_{dip}$ leads to different regimes of scattering as a function of the molecular dipole moment and the interlayer spacing. Namely, bound molecular states or BCS pairing can occur. Below we discuss the region of the parameters at which bound states can be formed.
\begin{figure}[htbp]
\begin{center}
\includegraphics[width=3.5in]{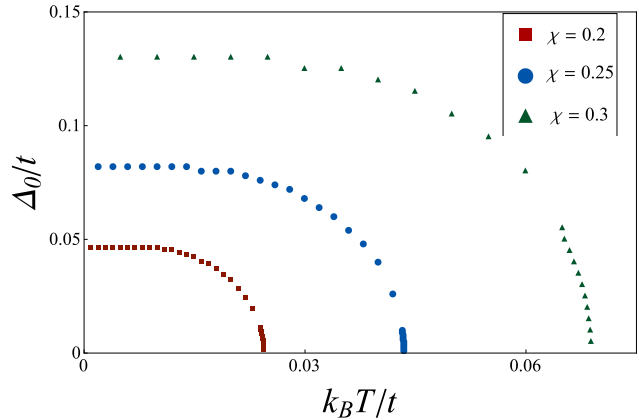} 
\end{center}
\caption{(Color online) Energy gap vs thermal energy (Eq. \eqref{gap}) for $\Lambda=0.5$. Calculations were done for square lattices of $N_x=N_y=120$ and half filling occupation.}
\label{Fig2}
\end{figure}

\begin{figure}[htbp]
\begin{center}
\includegraphics[width=3.5in]{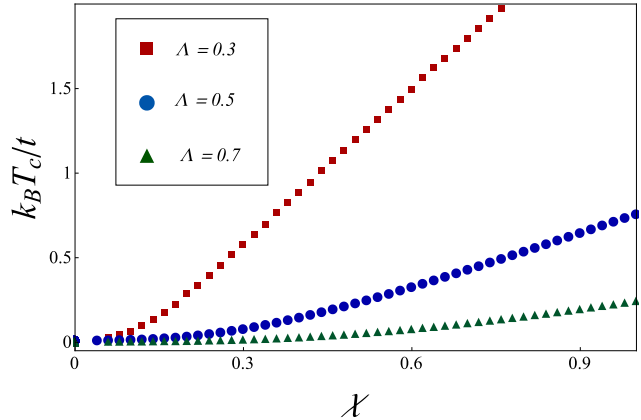} 
\end{center}
\caption{(Color online) Critical temperature as a function of the dimensionless interaction strength $\chi$ for $\Lambda=0.5, \, 0.75\, ,1.0$.}
\label{Fig3}
\end{figure}

To determine the critical temperature as a function of the parameters described above, we assume half-filling occupation. For all of our numerical calculations we work within the first Brillouin zone ($-\frac{\pi}{a} \le k_x,k_y \le \frac{\pi}{a}$) and use lattices of size $N_x=N_y=120$, which produce the same quantitative results than bigger lattices. The chosen values for molecular dipole moments and masses correspond to those  used in recent experiments on ultracold polar molecules \cite{Kaden}. In Fig. (\ref{Fig2}), we can observe the critical temperature at which the superfluid state occurs, within the BCS regime, for different values of the coupling strength $\chi$ and fixed $\Lambda =0.5$. Further, in Fig. \ref{Fig3} we plot the BCS critical temperature $T_c$ versus $\chi$ for different values of $\Lambda$, in order to illustrate the dependence of the critical temperature on those parameters.

{\it Ginzburg-Landau Free-Energy.-} We now expand the effective action, given by Eq. (\ref{Seff}), in terms of the new bosonic field $\Delta$ around the phase transition, and relate it to the Ginzburg-Landau Free-Energy \cite{Altland}
\begin{equation}
S_{eff}[\Delta,\Delta^*]=\hbar\beta F_{L}(\Delta).
\end{equation}
To do this, we write $\mathbb{G}_\Delta^{-1}(k,q)= \mathbb{G}_0^{-1}(k) + \Sigma_\Delta (k,q)$, where $\mathbb{G}_0^{-1}(k) = i\hbar \omega_n\mathbb{I}+ \sigma_z (\epsilon_k-\mu)$ and $\Sigma_\Delta (k,q)= \sigma_x \Delta(k,q)$, thus $\mathbb{G}_\Delta^{-1}(k,q)= \mathbb{G}_0^{-1}(k)(\mathbb{I}+ \mathbb{G}_0(k) \Sigma_{\Delta}(k,q))$. From the expansion $\ln \mathbb{G}^{-1}(k,q)= \text{ln} \mathbb{G}_0^{-1}(k)+\sum_{m=1}^\infty \frac{1}{m} \left( \mathbb{G}_0(k) \Sigma_{\Delta} (k,q) \right)^m$ we verify that odd powers of $ G_0 \Sigma$ do not contribute to the trace in $S_{eff}[\Delta]$, while even powers are explicitly given by 
 
$$
\left( \mathbb{G}_0 \Sigma_\Delta \right)^{2m} = \mathbb{I} \frac{ |\Delta(k)|^2 \left ( G_{11}^0 G_{22}^0 \right)^m}{\left [ (\hbar \omega_n)^2 +(\epsilon_{\vec{k}}-\mu)^2 \right]^{2m}},$$
being $G_{11}^0= -i \hbar \omega_n + (\epsilon_{\vec{k}}-\mu)$ and $G_{22}^0= -i \hbar \omega_n - (\epsilon_{\vec{k}}-\mu)$. 
As usual within the GL theory and well justified near the pairing transition temperature ($\Delta$ small), we truncate the expansion up to quartic order. This procedure leads us to determine the coefficients of the free energy $F = a(T) \Delta^2+ b(T) \Delta^4$.  As expected, the coefficient  $a(T)$ bears the same information as that given by the gap in Eq. (\ref{gap}), while $b(T)$ conveys the information on the stability of the equilibrium state. We emphasize that within the mean field approach, the whole information about the dipole interaction is contained in the critical temperature $T_c$. The coefficient $b(T)$ is found to be, 
\begin{equation}
 b(T)= \frac{1}{\Omega}  \sum_{\vec{k}}\left( -\beta^2 \frac{\sech ^2 \frac{(\epsilon_{\vec{k}}- \mu) \beta}  {2} }{8(\epsilon_{\vec{k}}-\mu)^2} + \frac{1}{4} \beta\frac{ \sinh \frac{(\epsilon_{\vec{k}}- \mu) \beta}  {2} } {(\epsilon_{\vec{k}} -\mu)^3}\right).
\end{equation}
 It is important to emphasize that, while the expression for $b(T)$ appears to be independent of the 2D lattice confinement and of the two-molecule interaction potential, their influence is exhibited through the critical temperature $T_c$, which it has been shown above to, in turn, strongly depend on the dipolar potential, namely on $\chi$ and $\Lambda$. As expected, the coefficient $b(T_c)$ is always positive.

{\it Superfluid density.-} To determine the superfluid density we also work within the perturbative scheme, neglecting quantum fluctuations, and following the functional Ginzburg-Landau free-energy theory for second order phase transitions \cite{Roth, Martikainen,Liu,Paananen}. That is, we expand the free energy in terms of the order parameter $\Delta$, incorporating a phase $\Delta= \Delta_0 e^{i \vec{\Theta} \cdot {\vec R_i}}$, where $\vec{\Theta}=(\frac{\Theta_x}{N_x a},\frac{\Theta_y}{N_y a})$. This procedure allows us to separate the density into the normal and superfluid components. In order to make explicit the dependence on the phase parameter $\vec{\Theta}$ and using the symmetry associated to the conservation of total molecule number, a gauge transformation can be performed on the operators that create and annihilate particles at the lattice sites in both layers: $( \hat{A}^\dagger _i,  \hat{A}_i)$, $( \hat{B}^\dagger _i,  \hat{B}_i)$ $\rightarrow$ $( \hat{A}^\dagger _i e^{i \vec{ \Theta} \cdot {\vec R_i}}, \hat{ A}_i  e^{-\vec{\Theta} \cdot {\vec R_i}})$, $( \hat{B}^\dagger _i e^{i  \vec{\Theta} \cdot {\vec R_i}},  \hat{B}_i  e^{-\vec{\Theta} \cdot {\vec R_i}})$. By performing this transformation in Hamiltonian given by Eq. (\ref{Hk}), having written it in its lattice representation \cite{HLattice}, one ends with the following expression for the Hamiltonian (\ref{Hk}):
\begin{eqnarray}
 \hat{H}_{\Theta}&=&  \hat{H}_0+ \sum_{{\vec k}} ( \hat{a}^{\dagger}_{\vec{k}} \hat{ a}_{\vec{k} }+ \hat{b}^{\dagger}_{\vec{k} }  \hat{b}_{\vec{k} })  \sum_{\alpha=x,y} \left(\Theta_{\alpha} \frac{\partial \epsilon_{ \vec k}  }{\partial k_{\alpha} } + \frac{\Theta^2_{\alpha}}{2} \frac{\partial^2 \epsilon_{ \vec k}  }{\partial k_{\alpha}^2 }\right)+ \hat{H}_I,
\label{Hk}
\end{eqnarray}
that is, associated to the phase $\vec{\Theta} \cdot {\vec R}_i$, there appears a contribution to the kinetic energy. Thus, in this scheme, the superfluid density component can be calculated as \cite{Paananen}
\begin{equation}
\rho_{\alpha,\alpha'}=  \lim_{\Theta \rightarrow 0} \frac{1}{Nt}\frac{ F_{\Theta} -F_0}{ \Theta_{\alpha} \Theta_{\alpha'} }=\frac{1}{Nt}\frac{\partial^2 F_{\Theta}}{\partial \Theta_{\alpha} \partial \Theta_{\alpha'}}, \, \alpha=\{x,y\}, \label{ro}
\end{equation}
being $F_0$ and $F_{\Theta}$ the free energies of the normal and superfluid phases respectively. The determination of the superfluid density $\rho_{\alpha,\alpha'}$, using Eq. (\ref{ro}), can be straightforwardly accomplished by noting that in the effective action, see Eq. (\ref{Seff}), an extra term appears associated to the phase of the gap. Thus, one can write, in an explicit way: $\mathbb{G}_{\Delta}^{-1}(k) \rightarrow \mathbb{G}_{\Theta}^{-1}(k)$ and consider perturbations up to second order. That is,  keeping quadratic terms of $\Theta$ in $\mathbb{G}_{\Theta}^{-1}(k)= \mathbb{G}_{\Theta=0}^{-1}(k) \left(\mathbb{I}- \mathbb{G}_{\Theta=0}(k) \Sigma_{\Theta}(k) \right)$. The final expression for $\rho_{\alpha,\alpha'}$ is
\begin{equation}
\rho_{\alpha\alpha'}=\frac{1}{\Omega a^2}\sum_{\vec{k}}\left(n(\vec{k})\frac{\partial^2\epsilon_{\vec{k}}}{\partial k_{\alpha}\partial k_{\alpha'}}-Y(\vec{k})\frac{\partial \epsilon_{\vec{k}}}{\partial k_\alpha}\frac{\partial \epsilon_{\vec{k}}}{\partial k_{\alpha'}}\right),
\end{equation}
where $n(\vec{k})$ is the momentum distribution and $Y(\vec{k})$ is the Yoshida distribution defined as $Y(\vec{k})^=\beta\sech^2(\beta\xi_{\vec{k}}/2)/4$, with $\xi_{\vec k}= \sqrt{(\epsilon_{\vec k}-\mu)^2 + \Delta^2}$. Since the off-diagonal terms of the superfluid tensor are small $\rho=(\rho_{xx}+\rho_{yy})/2$.
\begin{figure}[htbp]
\begin{center}
\includegraphics[width=3.5in]{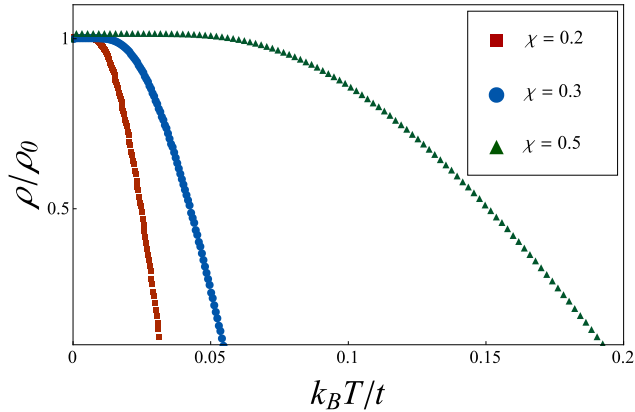} 
\end{center}
\caption{(Color online) Superfluid density for different values of the dimensionless interaction strength $\chi$ and $\Lambda=0.5$.}
\label{Fig4}
\end{figure}
Figure \ref{Fig4} shows the superfluid density fraction as a function of the temperature for three different values of the interaction strength. We denote by $\rho_{0}$  the superfluid density fraction at $T=0$. As it is well known the superfluid density fraction at zero temperature is different from unity in the lattice \cite{Paananen}.

{\it Berezinskii-Kosterlitz-Thouless temperature.-} When the interaction $\chi$ is strong enough, the predicted critical temperature becomes an artifact of the approximation since the BCS approach fails. To determine the relationship of the critical temperature with the coupling interaction, the presence of bounded pairs should be included. This procedure is delineated in Ref. \cite{SadeMelo} and gives rise to a clear identification of the Berezinskii-Kosterlitz-Thouless (BKT) transition temperature, as that at which a fraction of bound pairs dissociate. Thus, to estimate the region at which our treatment is valid, we compare it with the mentioned procedure. This  yields the following result,
\begin{equation}
k_{B}T_{BKT}=\frac{\pi \rho}{8 m_{eff}}. \label{BKT}
\end{equation}
\begin{figure}[htbp]
\begin{center}
\includegraphics[width=3.5in]{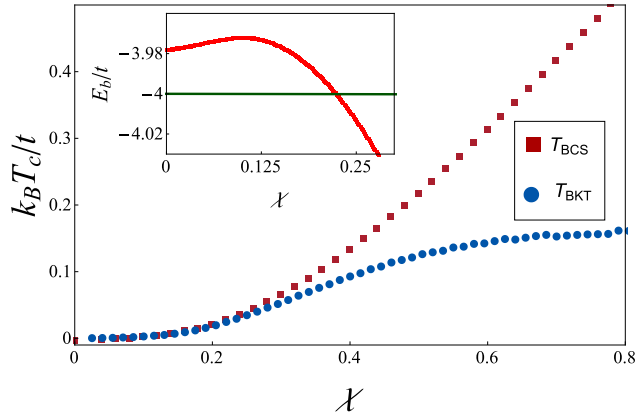} 
\end{center}
\caption{(Color online) BCS critical temperature $T_c$ and the Berezinskii-Kosterlitz-Thouless $T_{BKT}$ transition temperature as a function of the interaction strength $\chi$, for $\Lambda=0.5$. The inset correspond to the binding energy of the two-molecule system in the presence of the lattice.}
\label{Fig5}
\end{figure}
In Fig. \ref{Fig5} we plot the critical temperature as a function of the dimensionless interaction strength $\chi$ for both, our BCS scheme and the incorporation of bounded pairs, see Eq. (\ref{BKT}). We observe how in the weakly interacting regime the temperature $T_{BKT}$ is close to $T_{BCS}$, while they separate in the strong interaction region $\chi \gtrsim 0.3$. Beyond this value, the critical temperature $T_{BKT}$ corresponds to a phase in which dimerization dominates \cite{Gadsbolle}. It is important to note that while in the homogeneous system the mass is fixed, here, the presence of the lattice determines the effective mass, as a tunable parameter, that yields the scale at which the transition from BCS phase to dimerized BEC phase occurs. The inset of Fig. \ref{Fig5} shows the pair binding energy using a variational wave function for the two-molecule problem, $\phi (r,\gamma)= e^{-\gamma r}$ with $\gamma$ a variational parameter, within the presence of the 2D lattices.

The system of dipolar Fermi molecules in a 2D square lattice here proposed can be realized as an application of the recent experiments with ultracold molecules with anisotropic interactions. In particular, those with with polar molecules $^{40}\text{K} ^{87}\text{Rb}$ confined in optical lattices, of a lattice constant of $520$ nm  approximately \cite{Kaden}. For molecules such as $^{40}\text{K}^{87}\text{Rb}$ and $^6\text{Li}^{133}\text{Cs}$ the dipole strength $d$ can be varied with an external electric field from $0.2$ to $1.5$ D. Those experiments allow us to suggest typical interlayer spacings of $\lambda$ between $200-500$ nm, well within current experimental limits.

In conclusion, we have studied superfluidity in polar Fermi molecules placed in 2D square optical lattices bilayers. We have used a functional integral formulation to study the finite temperature BCS state as a function of the coupling strength $\chi$ and the dimensionless parameter that measures the scaling between the interlayer separation and the lattice size. To determine the superfluid density we employed the Ginzburg Landau free energy scheme by exploiting the symmetry associated to the isotropy in the lattice and the total molecule number conservation. Given the current experimental context, our study constitutes a strategy for handling the critical temperature of the superfluid state by changing both the dipole strength $d$ and the interlayer spacing among the 2D square optical lattices.

{\it Acknowledgments}

\noindent
This work was partially funded by grants IN107014 DGAPA (UNAM) and LN-232652 (CONACYT). A.C.G. acknowledges a scholarship from CONACYT.

\end{document}